\documentclass[12pt,letterpaper,aps,manuscript]{revtex}
\usepackage[T1]{fontenc}
\usepackage{times}
\usepackage{graphics}

\makeatletter

\providecommand{\LyX}{L\kern-.1667em\lower.25em\hbox{Y}\kern-.125emX\@}

\makeatother

\begin{document}

\draft

\title{Determination of the \protect\( \pi \protect \)NN Coupling Constant in the
VPI/GW \protect\( \pi \protect \)N\protect\( \rightarrow \protect \)\protect\( \pi \protect \)N
Partial-Wave and Dispersion Relation Analysis}

\author{M.M. Pavan\thanks{
email: mpavan@mit.edu
} }

\address{\emph{Lab for Nuclear Science, M.I.T., Cambridge, MA 02139}}

\author{R.A. Arndt }

\address{\emph{Dept. of Physics, Virginia Polytechnic and State University, Blacksburg,
VA 24061} }

\author{I.I. Strakovsky and R.L. Workman }

\address{\emph{Dept. of Physics, The George Washington University, Washington, D.C.
20052}}

\date{\today{}}

\maketitle
\begin{abstract}
Our extraction of the pion-nucleon coupling constant from \( \pi  \)p elastic
scattering data is outlined. A partial wave analysis (\( T_{\pi }<2100 \) MeV)
is performed simultaneously with a fixed-\( t \) dispersion relation analysis
(\( T_{\pi }< \) 800 MeV). The \( \pi NN \) coupling constant \( g^{2}/4\pi  \)
is searched to find the best fit. The result 13.73\( \pm  \)0.01\( \pm  \)0.07
(first error statistical, second systematic) is found to be insensitive to database
changes and Coulomb barrier corrections. This value satisfies important elements
of low energy QCD like the Goldberger-Treiman discrepancy, the Dashen-Weinstein
sum rule, and chiral perturbation theory predictions of threshold pion photoproduction.\\
\end{abstract}

\section{Introduction}

In recent times, the pion nucleon coupling constant \( g^{2}/4\pi  \) has remained
an illusive parameter. Many groups have worked on the problem using a number
of different approaches, yet there is still no consensus on its value to the
1\% level. This coupling is a very important \emph{input} parameter in low energy
QCD and nuclear physics, hence one requires its value to be determined as precisely
as possible. In one example, the Goldberger-Treiman relation \cite{gol58} connects
the \( \pi NN \) coupling constant to the well-known weak interaction and hadronic
quantities. The deviation in the relation has important implications \cite{dom85}
in low energy QCD. In another example, the Dashen-Weinstein sum rule \cite{dash69}
connects the coupling constant (via the Goldberger-Treiman discrepancy) to the
ratio of the strange and light quark masses. Using this relation and the ''textbook''
value \( g^{2}/4\pi  \) =14.3, it has been argued that the large quark condensate
assumption of standard chiral perturbation theory may not be valid so that a
``generalized'' form of the theory may be required \cite{fuch90}. From only
these two examples, clearly it is of fundamental importance to pin down \( g^{2}/4\pi  \)
.

Our recent partial wave analyses of \( \pi N \) scattering data up to \( T_{\pi }=2100 \)
MeV (latest publication in \cite{arndt95}) have included constraints from a
simultaneous fixed-\( t \) dispersion relation analysis. Our most recent (preliminary)
solution (SM99 \cite{sm99}) adds the forward ``derivative'' \( E^{\pm } \)
dispersion relations and the fixed-\( t \) \( C^{\pm } \) dispersion relation
to our suite of forward \( C^{\pm } \) and fixed-\( t \) \( B_{\pm } \) (``H\"uper'')
dispersion relations. In these dispersion relations, the coupling constant \( g^{2} \)
is an \emph{a priori} unknown parameter. The nucleon pole (Born) term is a well
defined quantity, and extracting the coupling constant from the dispersion relation
does not involve extrapolations or interpolations. The benefits of using fixed-\( t \)
dispersion relations to obtain \( \pi N \) scattering parameters like the coupling
constant have been extensively discussed by H\"ohler in \cite{hoehler83}. 

In their influential analysis, Bugg, Carter, and Carter \cite{bugg73} employed
the fixed-\( t \) \( B^{+} \) dispersion relation (see Eqn. \ref{eqn:hamDR})
and their own partial wave analysis over a narrow energy range (110-280 MeV)
to extract the coupling, obtaining \( g^{2}/4\pi  \) =14.3\( \pm  \)0.2. The
analysis was not \emph{constrained} to satisfy this dispersion relation, so
that value is simply what came out of the data via the partial wave analysis.
This value was subsequently used in the Karlsruhe analyses, in particular in
the fixed-\( t \) dispersion relation analysis of Pietarinen \cite{piet76}
used to constrain the partial wave solution ``KH80'' \cite{koch80} from which
the same coupling constant (14.3) was ``extracted'' using the so-called ``H\"uper''
dispersion relation. To our knowledge, no analyses were performed to test whether
other values of \( g^{2}/4\pi  \) gave better results i.e. better satisfied
dispersion relations, or better fits to data.

In the following, we outline our approach to extracting the pion nucleon coupling
constant from the \( \pi N \) scattering data, where unlike previous analyses
involving dispersion relations, it is treated as a free parameter to be determined
by \( \chi ^{2} \) minimization. We discuss briefly the method in Section \ref{sec:disprel},
followed by the results in Section \ref{sec:results} and some important systematic
checks we've made in Section \ref{sec:systchecks}. We close in Section \ref{sec:summary}
with a summary and some conclusions.

\section{Dispersion Relations and the Coupling Constant}

\label{sec:disprel}

The multi-energy partial wave analysis part of our analysis procedure has been
described in Refs.\cite{vpiSolns} . This part of the analysis has \emph{no}
explicit dependence on the coupling constant \( g^{2}/4\pi  \) and so will
not be discussed here. The interested reader is referred to those publications.

The sensitivity to the coupling constant enters through the fixed-\( t \) dispersion
relations used as constraints to the partial wave analysis. We employ the forward
(\( t \)=0) subtracted \( C^{\pm }(\omega ) \), and derivative \( E^{\pm }(\omega ) \)
dispersion relations, as well as the unsubtracted fixed-\( t \) H\"uper (\( B_{\pm }(\nu ,t) \)
) and \( C^{+}(\nu ,\, t) \) dispersion relations. All are implemented from
\( 20<T_{\pi }<800 \) MeV, and the fixed-\( t \) relations are also applied
over \( -0.3\leq t\leq 0 \) GeV\( ^{2} \)/c\( ^{2} \). The isovector constraints
can be applied to even higher energies, but presently we are having difficulty
satisfying the isoscalar \( C^{+} \) and \( E^{+} \) dispersion relations
at the higher energies. Since above 2100 MeV we employ the imaginary parts from
the KH80 analysis \cite{koch80}, and this solution also does not satisfy those
isoscalar dispersion relations very well around 1 GeV, the incompatability of
our solution and KH80 may be at the root of this problem. This is still under
investigation. The energy range over which the constraints limits the covered
momentum transfer range to \( t>-0.3 \) GeV\( ^{2} \)/c\( ^{2} \), which
is safely within the region where the partial wave expansion converges \cite{hoehler83}.
Nonetheless, the kinematic range over which the constraints are applied is entirely
sufficient for determinations of the ccoupling constant since it spans the \( \Delta  \)
resonance region which dominates the dispersion integrals, and where there are
the most abundant and precise data sets.

\begin{figure}
{\par\centering \resizebox*{0.9\columnwidth}{!}{\includegraphics{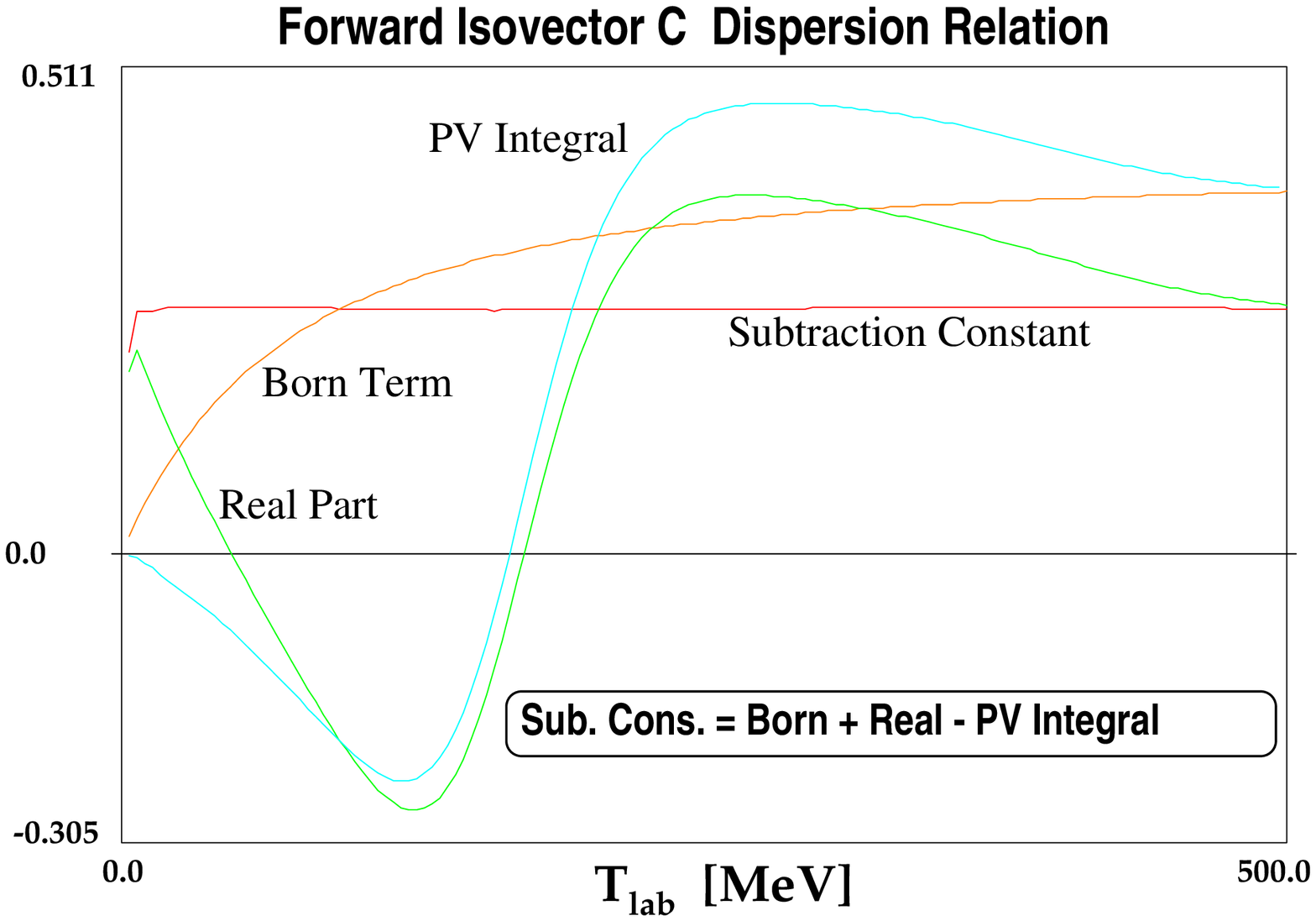}} \par}

\caption{\label{fig:fwdC-}The forward isovector \protect\( C^{-}(\omega )\protect \)
dispersion relation broken down into its component parts: the Born (nucleon
pole) term, the real part of C, the subtraction constant (``Sub. Cons.'')
and the principle value (``PV'') integral over a function of the imaginary
part of C (refer to Eqn.\protect\ref{eqn:fwdC-DR}). The dispersion relation
is satisfied if the subtraction constant is independent of energy. In our analysis,
at several kinematical points the squared difference of the real part and the
value required to ``flatten'' the constant is introduced as a \protect\( \chi ^{2}\protect \)
penalty function to be minimized. This ensures that once the solution converges,
the dispersion relation is satisfied to some prescribed accuracy (\protect\( \sim 1\%).\protect \)
This procedure is followed for all the dispersion relations used in our analysis.}
\end{figure}

In general, a fixed-\( t \) dispersion relation relates the real part of the
amplitude at some energy to a principal value integral over the imaginary parts
at \emph{all} energies, plus a nucleon pole contribution (Born term), and in
the case of subtracted relations, an additional \emph{subtraction constant}
which is in general energy \emph{independent}. (For the definitive discussion
of dispersion relations, see Ref. \cite{hoehler83}.) This form is demonstrated
by the special case of the forward \( C^{-}(\omega ) \) (isovector) dispersion
relation in Eqn.\ref{eqn:fwdC-DR} :

{\par\centering 
\begin{equation}
\label{eqn:fwdC-DR}
Re\, C^{-}(\omega )=C^{-}(\mu )+(1-\frac{\omega }{\mu })C^{-}_{N}(\omega )+\frac{2(\omega ^{2}-\mu ^{2})}{\pi }\int ^{\infty }_{\mu }\frac{d\omega '\, \omega \, Im\, C^{-}(\omega ')}{(\omega '^{2}-\omega ^{2})(\omega '^{2}-\mu ^{2})}
\end{equation}
\par}

where \( \mu ,\, M \) are the charged pion and proton masses, \( \omega =T_{\pi }+\mu  \)
is the total pion energy, \( \omega _{B}=\frac{\mu ^{2}}{2M} \) is the (unphysical)
pion energy at the nucleon pole, \( C^{-}_{N}(\omega )=\frac{\omega \, \omega _{B}}{\omega ^{2}_{B}-\omega ^{2}}\frac{g^{2}}{M} \)
is the Born term, and \( C^{-}(\mu )=4\pi \, a_{0+}^{-}(1+\mu /M) \) is the
subtraction constant, with the isovector s-wave scattering length \( a^{-}_{0+} \).
Refer to Fig. \ref{fig:fwdC-} for a graphical depiction of each term in Eqn.
\ref{eqn:fwdC-DR}.

\subsection{Implementation of Dispersion Relation Constraints.}

The \( \pi NN \) coupling constant (appearing in the Born term) and the subtraction
constants are \emph{a priori} unknown parameters . Our approach is to treat
them as \emph{searched parameters} to be determined by minimizing the \( \chi ^{2} \)
goodness-of-fit to the data \emph{and} the dispersion relations. For each dispersion
relation the constraint and its \( \chi ^{2} \) is evaluated as follows. First,
the coupling constant and all subtraction constants are fixed to some value.
With this set fixed, for each iteration in the partial wave analysis, the partial
waves are used to determine the principal value integral + Born term + subtraction
constant at a number of equally spaced kinematical points (\( \omega  \) for
forward dispersion relations, (\( \nu ,\, t) \) for fixed-\( t \)) to yield
a prediction for the real parts. The real part is then evaluated separately
using the partial waves. The difference ``\( Re \)(from PWA) - \( Re \)(from
DR)'' is then used to correct the real part of the partial waves for the next
iteration, and to calculate a \( \chi ^{2} \) using a desired accuracy (presently
set such that \( \chi ^{2}/pt.\sim 1 \) ) as the ``uncertainty''. This procedure
is iterated until the solution converges and a minimum overall \( \chi ^{2} \)
(fit to data + dispersion relations) is achieved. This yields the ``best''
solution corresponding a \emph{particular set} of dispersion relation parameters.
The entire analysis is repeated varying these parameters over a multi-dimensional
``grid'' and the minimum \( \chi ^{2} \) for each of these solutions recorded.
It is observed that e.g. the \( \chi ^{2} \) versus \( g^{2}/4\pi  \) curve
is a parabola, and so the parameters for the final solution are determined by
fitting quadratics (or bi-quadratics to 2-dimensional plots) to these curves
and selecting the parameters corresponding to the minimum.

This elaborate procedure has a number of benefits. From it one is able to define
a \emph{statistical} uncertainty for \( g^{2} \) by the variation which changes
the overall \( \chi ^{2} \) by 1. A \emph{systematic} uncertainty can be estimated
from the constancy of \( g^{2} \) over the applied kinematic range (``extraction
error''), and from the variation is the \( \chi ^{2} \) minimum for the separate
contributions of the dispersion relations and each of the three charge channels.

\subsection{\protect\( g^{2}/4\pi \protect \) from the H\"uper and \protect\( B^{+}(\nu \, t)\protect \)
Dispersion Relations}

When discussing the coupling constant, two dispersion relations merit special
consideration. One is the isoscalar \( B^{+}(\nu \, t) \) dispersion relation:

\begin{equation}
\label{eqn:hamDR}
\frac{g^{2}}{M}=\frac{\nu ^{2}_{B}-\nu ^{2}}{\nu }\left[ Re\, B^{+}(\nu ,\, t)-\frac{2\nu }{\pi }\int ^{\infty }_{\nu _{1}}d\nu '\frac{Im\, B^{+}(\nu ',t)}{\nu '^{2}-\nu ^{2}}\right] 
\end{equation}

where \( \nu =\omega +t/4M \) is the ``crossing'' energy variable, \( \nu _{B}=(t-2\mu ^{2})/4M \)
is the energy at the nucleon pole, and \( \nu _{1} \) is the threshold energy.
Written in this unsubtracted form, the coupling constant is the only unknown
parameter. Bugg, Carter, and Carter used this dispersion relation \cite{bugg73}
and obtained \( g^{2} \) by inputing in their phase shifts and then ``averaging''
over a kinematical range (\( \Delta \nu \, \Delta t \)) spanning the \( \Delta  \)
resonance. (Refer to Fig. \ref{fig:hamUXXX} for an example of this method in
our own analysis). This technique should yield a reliable estimate of \( g^{2} \)
since it is well known \cite{ham??} that the dispersion integral is dominated
by the first \( P_{33} \) (\( \Delta ) \) resonance, and evaluated there it
is satisfactorily convergent up to a few GeV (below which there still are abundant
data).

\begin{figure}
{\par\centering \resizebox*{0.9\columnwidth}{!}{\includegraphics{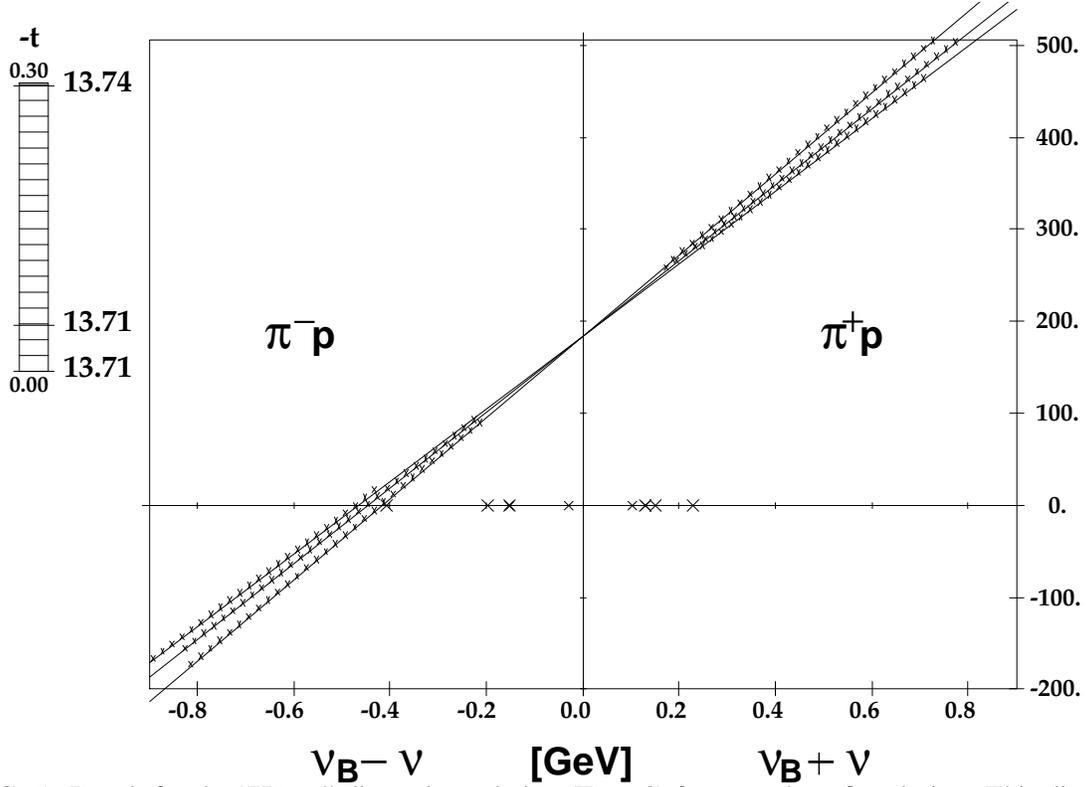}} \par}

\caption{\label{fig:hueperSM99}Result for the ``H\"uper'' dispersion relation (Eqn.
\protect\ref{eqn:hueperDR}) from our best-fit solution. This dispersion relation
is constructed so that the curves are linear, the y-intercept gives the coupling
\protect\( g^{2}/M\protect \), and the left (right)-hand side of the figure
is dominated by \protect\( \pi ^{-}p\protect \) (\protect\( \pi ^{+}p)\protect \)
data. This dispersion relation was used as one of the constraints, so by construction
there is virtually no t-dependence in the extracted couplings.}
\end{figure}

The other dispersion relation most useful for determining \( g^{2} \) is the
so-called ``H\"uper'' dispersion relation (see Refs. \cite{hoehler83} and
\cite{koch80}):

\begin{equation}
\label{eqn:hueperDR}
(\nu _{B}\pm \nu )\left[ \mp Re\, B_{\pm }(\nu ,\, t)\pm \frac{\nu }{\pi }\int _{\nu _{1}}^{\infty }\frac{d\nu '}{\nu }\left( \frac{Im\, B_{+}}{\nu '\mp \nu }+\frac{Im\, B_{-}}{\nu '\pm \nu }\right) \right] =\frac{g^{2}}{M}+\widetilde{B}(0,t)(\nu _{B}\pm \nu )
\end{equation}

where \( \widetilde{B}(0,t)=\frac{2}{\pi }\int ^{\infty }_{\nu _{1}}d\nu '\frac{Im\, B^{-}(\nu ',t)}{\nu '} \).
It is a clever combination of the invariant \( B \) amplitudes such that at
fixed \( t \), the left hand side of Eqn. \ref{eqn:hueperDR} is linear in
\( \nu _{B}\pm \nu  \) with the y intercept \emph{independent} of \( t \)
and equal to \( g^{2}/M \). This dispersion relation was used by Koch and Pietarinen
in their influential ``KH80'' analysis paper \cite{koch80}. (See Fig. \ref{fig:hueperSM99}
for an example from our analysis). Up to a few hundred MeV, the dispersion integrals
are dominated by the \( \Delta  \) resonance and suitably convergent up to
\( \sim  \) \emph{}4 GeV when evaluated there. They also have the property
that due to crossing symmetry, the ``left (\( \nu _{B}-\nu ) \) side'' is
dominated by \( \pi ^{-}p \) data, and the ``right \( (\nu _{B}+\nu ) \)
side'' by the \( \pi ^{+}p \) data. As will be shown in Section \ref{sec:coulcorrcheck},
one consequence is that extracting \( g^{2} \) from this dispersion relation
is relatively insensitive to the Coulomb corrections used in the partial wave
analysis.

\section{Results}

\label{sec:results}

\begin{figure}
{\par\centering \resizebox*{0.9\columnwidth}{!}{\includegraphics{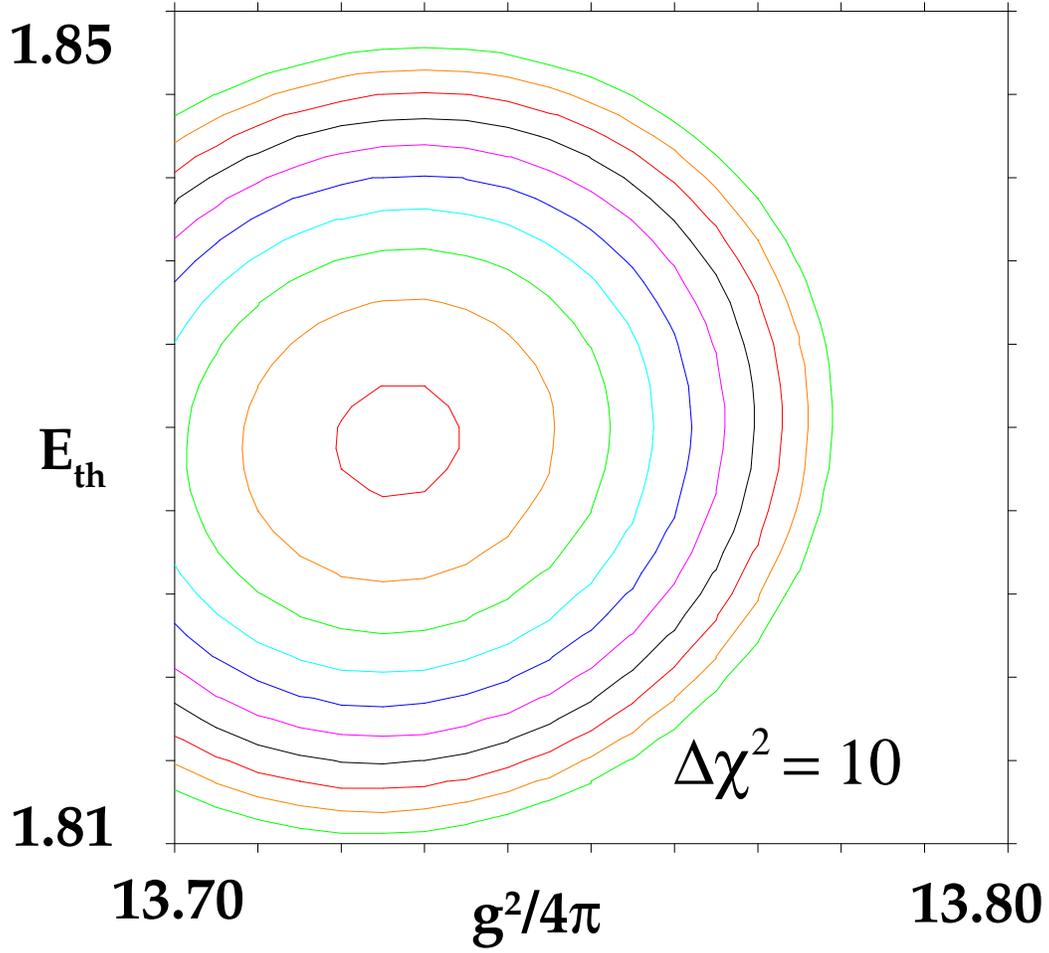}} \par}

\caption{\label{fig:g2vsEthChi2Map}Figure shows contours of constant total \protect\( \chi ^{2}\protect \)
in the (\protect\( E_{th},g^{2}/4\pi \protect \)) plane, where \protect\( E_{th}\protect \)
is the subtraction constant in the forward (derivative) \protect\( E^{+}\protect \)dispersion
relation. The contours were generated from a ``grid'' of 25 solutions where
(\protect\( E_{th},g^{2}/4\pi \protect \)) was fixed for each. A clear, deep
minimum is observed, as it is in general for the contour plots generated for
all pairs of dispersion relation parameters.}
\end{figure}

\begin{figure}
{\par\centering \resizebox*{0.9\columnwidth}{!}{\rotatebox{90}{\includegraphics{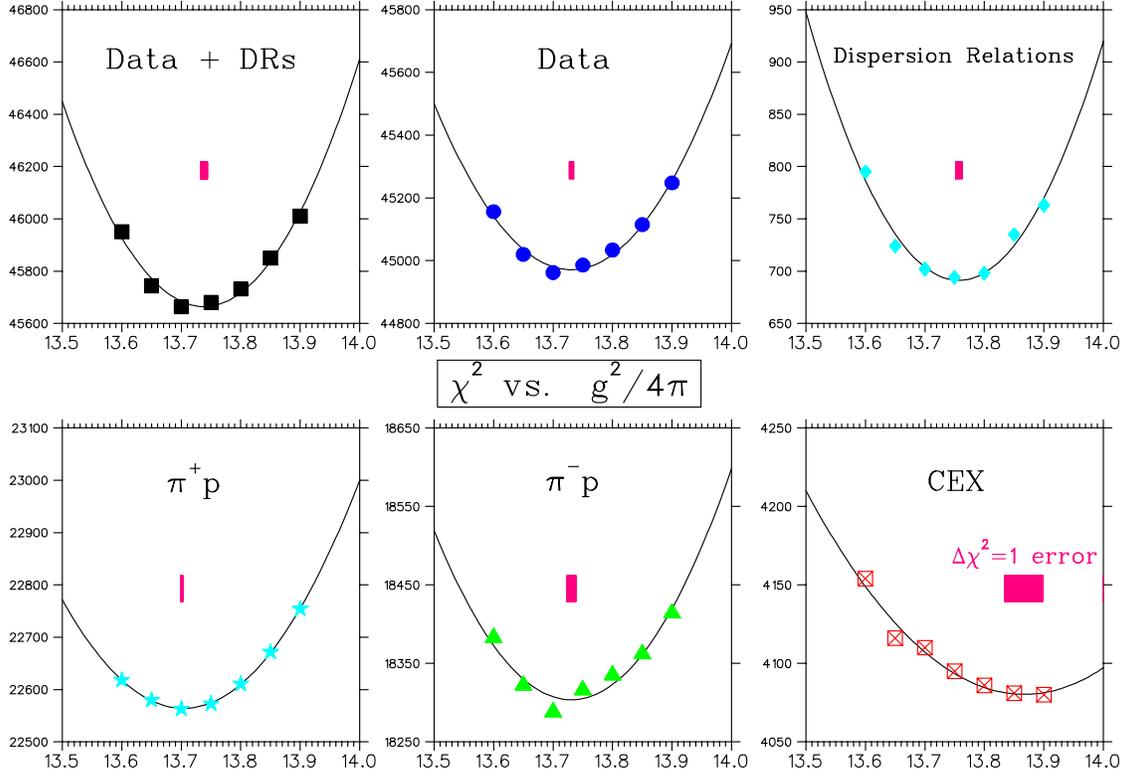}}} \par}

\caption{\label{fig:f2Chi2Breakdown}Best-fit \protect\( \chi ^{2}\protect \) as a
function of the coupling constant \protect\( g^{2}/4\pi \protect \), where
all other parameters were fixed to their optimal (best fit) values. Shown are
the total \protect\( \chi ^{2}\protect \) (``data+dispersion relation''),
and those for the dispersion relations, all data, and three charge channels
separately. Note that all curves minimize at very similar values (\protect\( \sim \protect \)13.73),
of which the (small) spread is one indication of the systematic uncertainty
in \protect\( g^{2}/4\pi \protect \). The bars indicate that \protect\( \Delta \chi ^{2}\protect \)=1
(statistical) uncertainty. }
\end{figure}

\begin{figure}
{\par\centering \resizebox*{0.9\columnwidth}{!}{\includegraphics{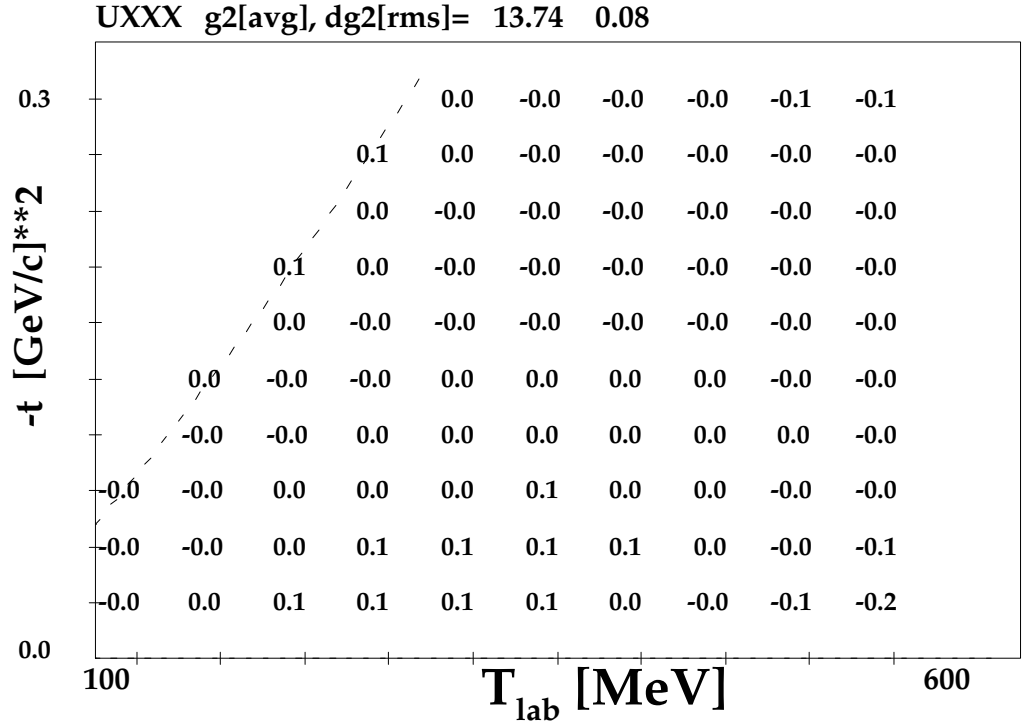}} \par}

\caption{\label{fig:hamUXXX}Results for the \protect\( B^{+}\protect \) dispersion
relation for a solution where NO dispersion relation constraints were used.
The extracted coupling 13.74\protect\( \pm \protect \)0.08 is very uniform
and perfectly consistent with the best fit value (see Fig.\protect\ref{fig:f2Chi2Breakdown}).
The consistency and uniformity demonstrates that the scattering data \emph{by
themselves} insist on the same coupling as the dispersion relations, and have
the expected analytic structure. This ``free'' solution satisfies most fixed-\protect\( t\protect \)
dispersion relations reasonably well (in particular the isovector ones, with
the sensitive isoscalar relations are less well satisfied), and yield similar
results to the constrained solution. This implies that the dispersion relation
constraints are merely ``fine tuning'' the partial wave amplitudes and not
forcing large changes.}
\end{figure}

Some \( g^{2} \) mapping results from our most recent solution (``SM99'')
\cite{sm99} are shown in figures \ref{fig:g2vsEthChi2Map} and \ref{fig:f2Chi2Breakdown}
. Figure \ref{fig:g2vsEthChi2Map} shows the two-dimensional constant \( \chi ^{2} \)
contours of the coupling constant versus the subtraction constant in the forward
derivative \( E^{+}(\omega ) \) dispersion relation. This contour plot was
generated using the results of 25 (5x5) solutions spanning the range shown where
all other dispersion relation parameters were fixed to their optimal values.
There is a distinct and deep minimum parabolic in shape with negligible correlation
between the two parameters. Fitting a bi-quadratic to the contours yields \( g^{2}/4\pi  \)
= 13.730\( \pm  \)0.009, where the uncertainty is statistical corresponding
the change \( \Delta \chi ^{2}=1 \). Very similar results are observed when
plotting constant \( \chi ^{2} \) contours of \( g^{2}/4\pi  \) versus the
scattering lengths.

The statistical uncertainty derived from the \( \chi ^{2} \) contour mappings
is clearly much smaller than the overall uncertainty. One way to estimate the
systematic uncertainty is to plot the one-dimensional curves of \( \chi ^{2} \)
versus \( g^{2}/4\pi  \) for each of the charge channels, their sum, and the
dispersion relation contributions separately, keeping all other parameters fixed
to their optimal values. This is shown in Fig.\ref{fig:f2Chi2Breakdown}. One
sees that all charge channels minimize near 13.73 (13.70, 13.73, 13.88 for \( \pi ^{+},\pi ^{-}, \)and
CEX respectively) as well as the dispersion relation contribution (13.76). This
spread gives one indication of the systematic uncertainty.

\begin{figure}
{\par\centering \resizebox*{0.9\columnwidth}{!}{\includegraphics{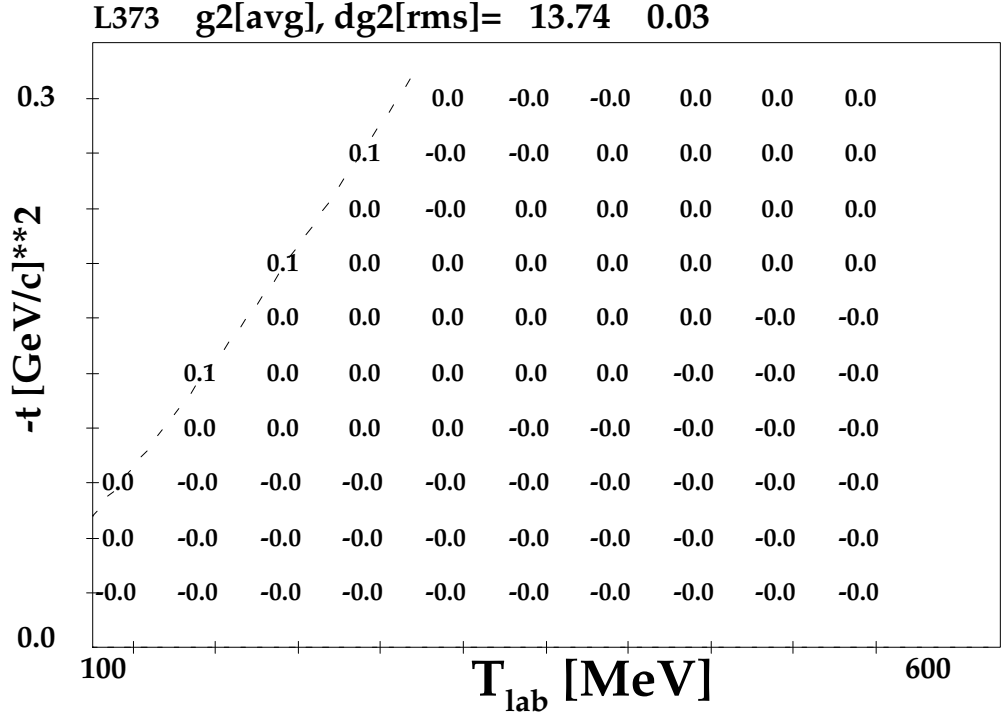}} \par}

\caption{\label{fig:hamSM99}Results for the \protect\( B^{+}\protect \) dispersion
relation (Eqn. \protect\ref{eqn:hamDR}) for a solution using the best-fit dispersion
relation parameters. The figure shows at each kinematical point \protect\( (T_{\pi ,}\, t)\protect \)
the deviation from the overall average of the coupling constant \protect\( g^{2}/4\pi \protect \)
. The extracted coupling 13.74\protect\( \pm \protect \)0.03 is very uniform.
This dispersion relation was not one of the constraints used in the analysis.}
\end{figure}

Another indication of the systematic uncertainty comes from the values extracted
from the \( B^{+}(\nu \, t) \) and H\"uper dispersion relations. The dispersion
relations in the solution SM99 were constrained with the \( g^{2}/4\pi  \)
=13.73 derived from the \( \chi ^{2} \) mappings, but fluctuations with respect
to \( \nu  \) and \( t \) can arise. The results are shown in figures \ref{fig:hamSM99}
and \ref{fig:hueperSM99}. One sees an almost negligible energy and t-dependence
of only about \( \pm 0.03 \) (0.2\%) over the full constraint range up to 800
MeV. We refer to this as the extraction uncertainty, and see again that it is
small with respect to the variations seen in Fig.\ref{fig:f2Chi2Breakdown}.

\subsection{Systematic Checks}

\label{sec:systchecks}

\subsubsection{Solution with NO Dispersion Relation Constraints}

A number of checks were made in order to gauge other sources of systematic uncertainty.
One check was to generate a partial wave analysis solution with \emph{no dispersion
relation constraints whatsoever} and then use the resulting amplitudes in the
dispersion relations to extract \( g^{2}/4\pi  \). This is in effect the approach
used by many other prior works, including that of Bugg, Carter, and Carter \cite{bugg73}.
This was done to see if the dispersion relation constraints were ``pulling''
\( g^{2} \) away from a value preferred by the data alone. This turns out \emph{not}
to be the case. Over the same (\( \nu ,\, t) \) ranged used in the constrained
analysis, the dispersion relation-free solution yields \( g^{2}/4\pi  \) =
13.66\( \pm 0.18 \) (1.3\%) from the \( B^{+}(\nu \, t) \) dispersion relation
(see Fig.\ref{fig:hamUXXX}), and 13.66\( \pm 0.07 \) from the H\"uper dispersion
relation (not shown), where only the extraction uncertainties are quoted. Up
to 450 MeV, where their are ample modern, precise cross section and polarization
data from the meson factories, these dispersion relations yield about 13.77\( \pm 0.07 \).
In fact up to \( \sim  \)500 MeV \emph{all} the fixed-\( t \) dispersion relations
used in the constrained analysis are reasonably well satisfied. This is an important
observation: to a decent approximation, the low and intermediate energy scattering
data exhibit the analytic properties expected of them, and so applying dispersion
relation constraints to a partial wave analysis solution has the effect of ``fine
tuning'' the amplitudes and not drastically altering them from their unconstrained
state.

\subsubsection{Coulomb Corrections}

\label{sec:coulcorrcheck}

The dispersion relations must use amplitudes from which all Coulomb contributions
have been removed (i.e. ``hadronic amplitudes''), consequently a systematic
contribution to \( g^{2} \) can enter through the Coulomb correction scheme
employed. The direct Coulomb and Coulomb phase rotation prescription used in
our analysis comes from the Nordita analysis \cite{trom77} which was used in
the Karlsruhe-Helsinki KH80 solution \cite{koch80}. Our prescription for the
Coulomb barrier correction has been criticized as being too simple (see e.g.
\cite{bugg93}). To test the effect of our Coulomb barrier correction scheme
on the coupling constant extraction, we made a solution taking the radical step
of \emph{ignoring it altogether.} The minimum in the overall \( \chi ^{2} \)
was found to be 13.70, varying from \( \sim  \)13.45 for \( \pi ^{+}p \) to
13.72 for \( \pi ^{-}p \) and 13.67 for charge exchange. Since surely our Coulomb
barrier correction is more accurate than ignoring it altogether, we conclude
that the systematic uncertainty due to our Coulomb barrier correction scheme
is not significant with respect to the other systematic uncertainties.

\begin{figure}
{\par\centering \resizebox*{0.9\columnwidth}{!}{\includegraphics{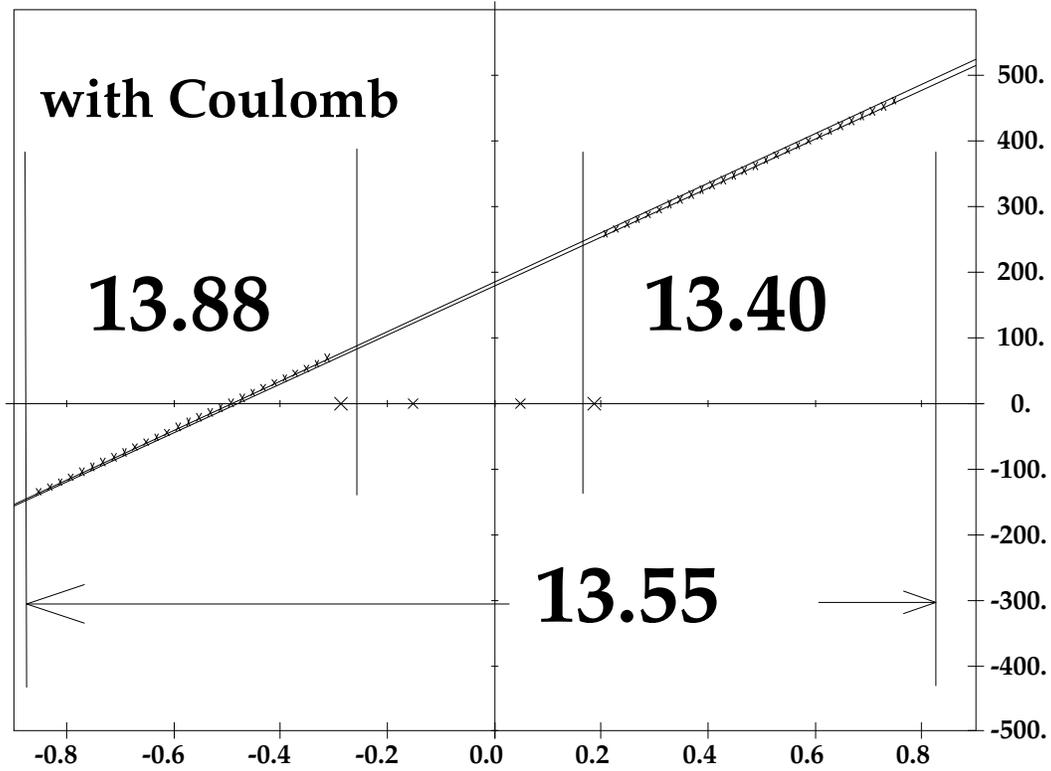}} \par}

\caption{\label{fig:hueperCoulCheck}H\"uper dispersion relation at \protect\( t=-0.15\protect \)
GeV\protect\( ^{2}\protect \)/c\protect\( ^{2}\protect \), but with the Coulomb
barrier correction \emph{reintroduced} into the amplitudes. Since the left side
is dominated by \protect\( \pi ^{-}p\protect \) data and the right side by
\protect\( \pi ^{+}p\protect \), one sees that the amplitudes are enhanced
on one side and suppressed on the other, such that the average remains relatively
constant (13.55 \emph{vs.} 13.73 = 1.3\% change) with respect to the normal
solution. This ``pivoting'' around the y-intercept implies that this dispersion
relation is \emph{insensitive} to the Coulomb barrier correction. Therefore
the difference between the Coulomb barrier correction used in our analysis and
a more sophisticated form will not cause a significant change in the coupling
constant.}
\end{figure}

It is interesting to note the effect of the Coulomb barrier correction on the
H\"uper dispersion relation. We reintroduced the correction into the hadronic
amplitudes and calculated the dispersion relation as before. The result is shown
in Fig.\ref{fig:hueperCoulCheck}. The correction suppresses (enhances) the
\( \pi ^{+}p \) (\( \pi ^{-}p) \) amplitudes. Since the plot is dominated
by \( \pi ^{+}p \) (\( \pi ^{-}p) \) data on the right (left) hand side, the
effect is to ``rotate'' the line around the intercept, leaving it, hence the
coupling constant, relatively unchanged. One sees in Fig.\ref{fig:hueperCoulCheck}
that the result changes by only 1.3\% (13.55 \emph{vs.} 13.73). Clearly this
insensitivity to the Coulomb correction makes the H\"uper dispersion relation
valuable for determining the coupling constant.

\subsubsection{Database Changes}

Another source of systematic uncertainty comes from the elastic pion proton
scattering database. As has been mentioned, it is well known that the \( \Delta  \)
resonance amplitude dominates most of the dispersion relations. It was shown
in the context of the Goldberger-Miyazawa-Oehme sum rule \cite{loch93} that
the difference between the Pedroni, et al. \cite{ped78} and Carter, et al.
\cite{car71}, total \( \pi ^{\pm }p \) cross sections results in a 1\% change
in \( g^{2} \), so it is important to study the database contribution uncertainties
in the context of our full analysis. 

We constructed a solution where the total cross section data of Pedroni, et
al., were removed, and one where the total cross section data of Carter, et
al., the total CEX reaction cross section data of Bugg, et al. \cite{bugg71},
and the differential \( \pi ^{\pm }p \) differential cross section data of
Bussey, et al. \cite{buss73} were removed. We found only a small change of
about \( \pm 0.04 \) with respect to our normal solution. As opposed to the
result in \cite{loch93}, the effect of a number of fixed-\( t \) dispersion
relation constraints and many data sets (differential, partial total, and polarization)
reduces the sensitivity to any single measurement

We constructed yet another solution where \emph{all} charge exchange data were
removed from the database. This CEX-less solution results in a best fit coupling
of 13.65, which is only 0.6\% lower than the nominal value 13.73. This is expected,
since one sees from Fig.\ref{fig:f2Chi2Breakdown} that the \( \chi ^{2} \)
minimum occurs at a larger \( g^{2}/4\pi  \) for CEX than for \( \pi ^{\pm }p \)
. 

It should be noted that in the process of analyzing the \( \pi ^{\pm }p \)
scattering data over the years, many new data have entered the database, and
we have tried many solutions accepting some and deleting other data sets. None
of these changes have ever caused a large change in \( g^{2}/4\pi  \) , and
the result has remained stable around 13.73 since at least 1993. We conclude
that the database incompatibilities that do exist in the current database do
not contribute greatly to the coupling constant uncertainty.

\section{Conclusion}

\label{sec:summary}

We have outlined our approach to extracting the pion nucleon coupling constant
\( g^{2}/4\pi  \) from the \( \pi p \) elastic scattering database using fixed-\( t \)
dispersion relations. Obtaining the coupling constant from these dispersion
relations is theoretically unambiguous and relies on the general principles
of analyticity, unitarity, and crossing symmetry \cite{hoehler83}. The coupling
constant was treated as a searched parameter to be determined by a least-squares
fit to the data and the dispersion relations. From our most recent analysis
(solution SM99), our result is \( g^{2}/4\pi  \) =13.73\( \pm 0.01\pm 0.07 \),
where the first uncertainty is statistical (corresponding to a change \( \Delta \chi ^{2}=1 \))
and the second is systematic. The latter uncertainty was estimated from the
differences in the \( \chi ^{2} \) minima for each of the charge channels and
the dispersion relations (Fig.\ref{fig:f2Chi2Breakdown}), from changes to the
\( \pi p \) scattering database, from modifications to the Coulomb barrier
corrections (Fig.\ref{fig:hueperCoulCheck}), from a solution with \emph{no}
dispersion relation constraints (Fig. \ref{fig:hamUXXX}), and from the variations
over the kinematical range where the dispersion relations were applied (Figs.
\ref{fig:hamSM99} and \ref{fig:hueperSM99}). 

Our result is in serious contradiction with the recent results from np backward
scattering differential cross section measurements (see e.g. \cite{eric95}
and other contributions to these proceedings) of \( g^{2}/4\pi \sim  \)14.5\( \pm 0.3 \),
though more in line with the results of the a similar analysis by Arndt, et
al. \cite{arndt99}. The difference is about 9 of our standard deviations. It
is very difficult to see how our result could be in error by this amount. Nevertheless
we are continuing our efforts to update the analysis as data new data come in,
and to refine our methods in order to extract the most precise value of \( g^{2}/4\pi  \)
possible.

It should be noted that the coupling constant result 13.73 resolves the problem
with the Goldberger-Treiman discrepancy (\cite{gol58},\cite{dom85}) which
is too large using the larger coupling. Also it has recently been shown \cite{goity99}
that the Dashen-Weinstein sum rule \cite{dash69} (corrected for some higher
order chiral effects) is not consistent with the larger coupling, favouring
instead a value near 13.7. It has also been shown \cite{g2photoprod} that recent
measurements of threshold pion photoproduction are more consistently described
in chiral perturbation theory with the smaller coupling than the larger. Along
with many other recent determinations which have arrived at a lower coupling
constant (refer to the review \cite{deswart97}, the fact that these important
aspects of low energy QCD are more consistently described with a coupling constant
near 13.7 rather than 14.3 is in our opinion a good argument in favour of adopting
the lower value as the current standard.

\section*{acknowledgments}

This work was supported in part by the U.S. Department of Energy Grants DE-FG02-99ER41110,
DE-FG02-97ER41038, and DE-FG02-95ER40901. MMP would like to thank T.E.O. Ericson
for his support while in Uppsala.

\newpage


\end{document}